\documentstyle[epsfig,preprint,aps]{revtex}
\begin{document} 
\draft
\preprint{\vbox{\hbox{IFT--P.011/98}
                \hbox{FTUV/97-60}
                \hbox{IFIC/97-90}
}}
%
%
\title{Constraints on Four Fermion Contact Interactions from
       Precise Electroweak Measurements} 
\author{M.\ C.\ Gonzalez--Garcia$^{1,2}$, A.\ Gusso$^1$ and S.\ F.\ Novaes$^1$}
\address{$^1$ Instituto de F\'{\i}sica Te\'orica, 
              Universidade  Estadual Paulista, \\  
              Rua Pamplona 145, CEP 01405--900 S\~ao Paulo, Brazil.\\
and \\
         $^2$ Instituto de F\'{\i}sica Corpuscular -- IFIC/CSIC \\
              Dept.\ de F\'{\i}sica Te\'orica, Universidad de Valencia \\
              46100 Burjassot, Valencia, Spain}
\date{\today}
\maketitle
\widetext

\begin{abstract}
We establish constraints on a general four--fermion contact
interaction from precise measurements of electroweak parameters.
We compute the one--loop contribution for the leptonic $Z$ width,
anomalous magnetic, weak--magnetic, electric and weak dipole moments of
leptons in order to extract bounds on the energy scale of these
effective interactions. 
\end{abstract}

\narrowtext
\section{Introduction}
Four--fermion contact interactions are able to describe the
dominant effect at low energies, arising from the existence of
quark and lepton substructure \cite{compo}. Such interactions can
be generated by the exchange of some common constituents between
the fermion currents or by the binding force that keeps the
constituents together. 

Recently, the phenomenology associated to these four--Fermi
contact interactions has become the subject of intense study  as
they have been proposed as a possible explanation
\cite{four,our:4} to the  high--$Q^2$ anomaly in the HERA
\cite{hera} data.  Both H1 and ZEUS experiments at HERA  have
reported the observation of an excess of events, compared with
the Standard Model prediction, in the reaction $e^+ p \to e^+ +
X$ at very high--$Q^2$. The H1 Collaboration observed events seem
to be concentrated at an invariant mass of $\sim 200$ GeV, what
could suggest the presence of a $s$--channel resonant state.  The
ZEUS Collaboration data, however, are more spread in invariant
mass.  The probability of a statistical fluctuation seems to be
quite small (less than $6 \times 10^{-3}$, for the H1 data).
Nevertheless, up to this moment, it is not possible to establish
the resonant or continuum aspect of the events although the
continuum aspect seems to be slightly favoured by  the most
recent data \cite{spiesberger}. Moreover, the existence of quark
substructure could also manifest as the enhancement of the
inclusive jet differential cross section at high $E_T$
\cite{tevatron}.

In this paper, we study the constraints on general four--fermion
contact interactions arising from the precise measurements of the
electroweak parameters. We evaluate the one--loop contribution of
the four-Fermi Lagrangian to the leptonic $Z$ width, anomalous
magnetic, weak--magnetic, electric and weak dipole moments of
leptons. The comparison with recent experimental data for these
parameters allow to extract bounds on the energy scale of these
effective interactions.

\section{One--Loop Contribution of the Four--Fermion Interaction}

We analyse the one--loop contribution of all possible
four--fermion contact interactions, {\it i.e.} represented by
scalar, vector and tensorial currents,
\begin{mathletters}
\label{lag:4}
\begin{equation}
{\cal L}_{\text{scalar}} =  \eta \frac{g^2}{\Lambda^2} 
\left[\bar{\psi}_m \left( V_S^m - i A_S^m \gamma_5 \right) \psi_m
\right] \; 
\left[\bar{\psi}_n \left( V_S^n - i A_S^n \gamma_5 \right) \psi_n \right]
\label{sca}      
\end{equation}
\begin{equation}
{\cal L}_{\text{vector}} = \eta \frac{g^2}{\Lambda^2} 
\left[\bar{\psi}_m \gamma^\mu \left( V_V^m - A_V^m \gamma_5 \right) \psi_m 
\right] \; 
\left[\bar{\psi}_n \gamma_\mu \left( V_V^n - A_V^n \gamma_5 \right) \psi_n 
\right]
\label{vec}       
\end{equation}
\begin{equation}
{\cal L}_{\text{tensor}} = \eta \frac{g^2}{\Lambda^2}
\left[\bar{\psi}_m \sigma^{\mu\nu} \left( V_T^m - i A_T^m \gamma_5 \right) 
\psi_m \right]\; 
\left[\bar{\psi}_n \sigma_{\mu\nu} \left( V_T^n - i A_T^n \gamma_5 \right) 
\psi_n \right]
\label{ten}
\end{equation}
\end{mathletters}
where $\Lambda$ is the energy scale of the effective interaction,
$m$ and $n$ are lepton and quark flavors, and $\eta = \pm 1$. In
general, these  Lagrangians do not respect the $SU(3)\otimes
SU(2)\otimes U(1)$ of the Standard  Model but solely respect the
$U(1)$ symmetry. In what follows we have assumed that the
fermionic currents do not mix flavors, unlike in the case of
effective operators that are generated at low energy by the
exchange of heavy leptoquarks. 

We write the matrix element of a neutral vector boson ($V =
\gamma, Z$) current in the general form: 
\begin{equation}
J^\mu  = e \; \bar{u}_f (p_1) \;  
\left( F_v \;  \gamma_\mu + 
       F_a \;  \gamma_\mu \gamma_5 +  
       F_m \;  \frac{i}{2 m_f} \sigma^{\mu\nu} q_\nu + 
       F_d \;  \frac{1}{2 m_f} \sigma^{\mu\nu} \gamma_5 q_\nu \right)
v_f (p_2)  \; , 
\label{current}
\end{equation}
where the form factors $F_i$, $i = v, \, a, \, m, \, d$, are
functions of $Q^2$, with $Q = p_1 + p_2$.  The form factors $F_v$
and $F_a$ are present at tree level in the Standard Model, {\it
e.g.\/} for the photon,
\begin{equation}
F_v^{\text{tree}} \equiv G_V = Q_f \;\; , \;\;\;\; 
F_a^{\text{tree}} \equiv G_A = 0
\label{g:a}
\end{equation}
where $Q_f$ is electric charge of the fermion in unities of the
proton charge, while for the $Z$ boson, 
\begin{equation}
F_v^{\text{tree}} \equiv G_V = \frac{1}{2 s_W c_W} 
(T_3^f - 2 \, Q_f \, s_W^2)
\;\; , \;\;\;\; 
F_a^{\text{tree}} \equiv G_A = \frac{1}{2 s_W c_W} \,  T_3^f
\label{g:z}
\end{equation}
where $s_W \, (c_W) = \sin \, (\cos) \theta_W$, and $T_3^f$ and
is the third component of the weak isospin of the fermion.  

The four--fermion interaction contribution to the form factors
$F_v$ and $F_a$ at one--loop level can alter the $Z$ boson width
to fermions. The form factor $F_m$ is responsible for an
additional contribution to the anomalous magnetic and
weak--magnetic moments of the fermion which are denoted by
$a^{\gamma}_f$ and $a^{Z}_f$, respectively,
\begin{eqnarray}
F_m (Q^2 = 0)       & = & a^{\gamma}_f \equiv \frac{(g_f - 2)}{2} \;,  
 \nonumber \\
F_m (Q^2 = M_{Z}^2) & = & a^{Z}_f \; . 
\end{eqnarray}

The form factor $F_d$ is related to electric ($d^e_f$) and weak
dipole moments  ($d^w_f$) by,
\begin{eqnarray}
\frac{e}{2 m_f} F_d (Q^2 = 0)     & = & d^e_f \;, \nonumber \\ 
\frac{e}{2 m_f} F_d (Q^2 = M_Z^2) & = & d^w_f \;.
\end{eqnarray}

\begin{figure}
\begin{center}
\mbox{\epsfig{file=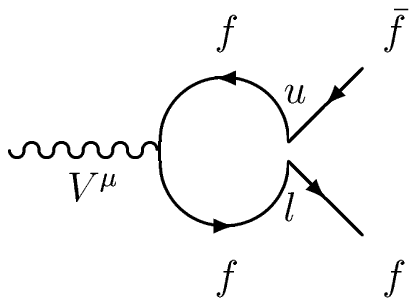,width=.6\linewidth}}
\end{center}
\caption{$s$--channel  diagram.}
\label{cs:fig}
\end{figure}

\begin{figure}
\begin{center}
\mbox{\epsfig{file=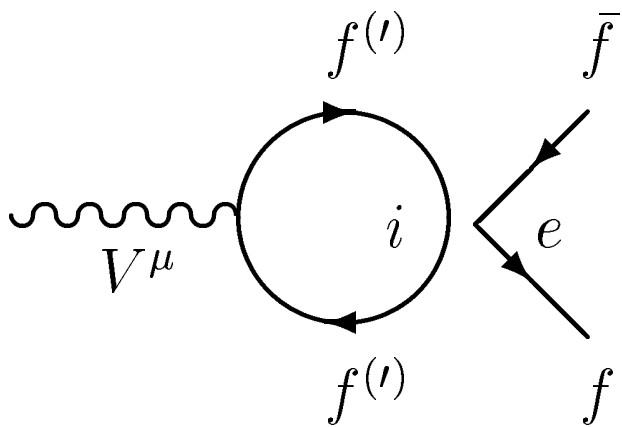,width=.6\linewidth}}
\end{center}
\caption{$t$--channel  diagram.}
\label{ct:fig}
\end{figure}

The contact interactions in Eq. (\ref{lag:4}) modify these form
factors at one-loop through the diagrams presented in Figs.\
(\ref{cs:fig}) and (\ref{ct:fig}). We obtain for the nonzero
one--loop contributions of the scalar Lagrangian (\ref{sca}) to
the $s$--channel diagrams,
\begin{eqnarray} 
F_v^S (s) &=& \eta \frac{g^2}{48 \pi^2 \Lambda^2} G_V  (V^u_S V^l_S +
 A^u_S A^l_S) Q^2
\log\left(\frac{\Lambda^2}{\mu^2}\right) \; ,
\nonumber \\
F_a^S (s) &=& -\eta \frac{g^2}{48 \pi^2 \Lambda^2} G_A (V^u_S V^l_S +
A^u_S A^l_S ) 
\left(6 M^2  - Q^2 \right) \log\left(\frac{\Lambda^2}{\mu^2}\right)\; ,
\nonumber \\
F_m^S (s) &=& \eta \frac{g^2}{8 \pi^2 \Lambda^2} G_V (V^u_S V^l_S - 
A^u_S A^l_S) M^2
\log\left(\frac{\Lambda^2}{\mu^2}\right) \; ,
\nonumber \\
F_d^S (s) &=& \eta \frac{g^2}{4 \pi^2 \Lambda^2} G_V (V^l_S A^u_S+
V^u_S A^l_S) M^2  
\log\left(\frac{\Lambda^2}{\mu^2}\right) \; ,
\label{sca:s}
\end{eqnarray}
where $M$ is the internal and external fermion mass, $G_{V,A}$ is
the vector (axial) coupling of the gauge boson, {\it c.f.} Eq.\
(\ref{g:a}) and (\ref{g:z}) and the indexes $u$ ($l$) denote the
constants associated  to the upper (lower) vertices of Fig.\ 
(\ref{cs:fig}). The scalar Lagrangian does not contribute to the
$t$--channel.

The contribution of interaction (\ref{vec}) to the $s$--channel
is,
\begin{eqnarray}
F_v^V (s) = \eta \frac{g^2}{48 \pi^2 \Lambda^2}  &&\Biggl\{
\left[6 G_A  M^2 - (G_V + G_A) Q^2\right] (V^l_V + A^l_V)(V^u_V + A^u_V)  
\nonumber \\
&& 
- \left[6 G_A  M^2 + (G_V - G_A) Q^2\right] (V^l_V - A^l_V)(V^u_V - A^u_V) 
\Biggr\}  
\log\left(\frac{\Lambda^2}{\mu^2}\right) \; ,
\nonumber \\
F_a^V (s) = - \eta \frac{g^2}{48 \pi^2 \Lambda^2}  &&\Biggl\{
\left[6 G_A M^2 - (G_V + G_A) Q^2 \right] (V^l_V + A^l_V)(V^u_V + A^u_V)
\nonumber \\
&&
+ \left[6 G_A M^2 + (G_V - G_A) Q^2 \right] (V^l_V - A^l_V)(V^u_V - A^u_V)  
\Biggr\} 
\log\left(\frac{\Lambda^2}{\mu^2}\right) \; .
\label{vec:s}
\end{eqnarray}
and to the $t$--channel,
\begin{eqnarray}
F_v^V (t) &=& \eta \frac{g^2}{12 \pi^2 \Lambda^2} V_{V}^e
\left[6 G_A^i A_{V}^i M_i^2 - (G_A^i A_{V}^i+ G_V^i V_{V}^i) Q^2 \right]
\log\left(\frac{\Lambda^2}{\mu^2}\right) \; ,
\nonumber \\
F_a^V (t) &=& -\eta \frac{g^2}{12 \pi^2 \Lambda^2} A_{V}^e 
\left[6 G_A^i A_{V}^i M_i^2 - (G_A^i A_{V}^i + G_V^i V_{V}^i) Q^2 \right]
\log\left(\frac{\Lambda^2}{\mu^2}\right) \; .
\label{vec:t}
\end{eqnarray}
where the index $i$ refers to the mass and coupling constants of
the internal fermion running in the loop and $e$ refers to the
external fermion ({\it c.f.\/} Fig. (\ref{ct:fig})).

Finally, the tensor Lagrangian (\ref{ten}) contributes to the
$s$--channel as,
\begin{eqnarray}
F_m^T (s) &=& -\eta \frac{g^2}{2 \pi^2 \Lambda^2} G_V (V^l_T V^u_T -
 A^l_T A^u_T)  M^2
\log\left(\frac{\Lambda^2}{\mu^2}\right) \; ,
\nonumber \\
F_d^T (s) &=& -\eta \frac{g^2}{\pi^2 \Lambda^2} G_V (V^l_T A^u_T+
V^u_T A^l_T)  M^2
\log\left(\frac{\Lambda^2}{\mu^2}\right) \; ,
\label{ten:s}
\end{eqnarray}
and to the $t$--channel as,
\begin{eqnarray}
F_m^T (t) &=& - \eta \frac{g^2}{\pi^2 \Lambda^2} G_V^i
(V_{T}^e V_{T}^i - A_{T}^e A_{T}^i)  M_e M_i
\log\left(\frac{\Lambda^2}{\mu^2}\right) \; ,
\nonumber \\
F_d^T (t) &=& - \eta \frac{g^2}{\pi^2 \Lambda^2} G_V^i 
(A_{T}^e V_{T}^i + A_{T}^i V_{T}^e)  M_e M_i
\log\left(\frac{\Lambda^2}{\mu^2}\right) \; .
\label{ten:t}
\end{eqnarray}

The loop contributions of the four--fermion interaction were
evaluated in $D = 4 - 2 \epsilon$ dimensions using the
dimensional regularization method \cite{reg:dim}, which is a
gauge--invariant regularization procedure. We have adopted the
unitary gauge to perform the calculations. The results in $D$
dimensions were obtained with the aid of the Mathematica package
FeynCalc \cite{feyn}, and the poles at $D = 4$ ($\epsilon=0$)
were identified with the logarithmic dependence on the scale
$\Lambda$. We retain only the leading non--analytical
contribution  from the loop diagram by making the identification
$2/(4-d) \rightarrow \log(\Lambda^2/\mu^2)$, where $\mu$ is the
scale involved in the process. At the $Z$ pole we use $\mu = M_Z$
and for processes involving real photons ($Q^2 = 0$) we choose
$\mu$ equal to the mass of the final state fermion. Notice that
the contributions to the photon form factor $F_v$ cancel at
$Q^2=0$ as required by the the QED Ward identities \cite{ward}.
Our bounds on the scale $\Lambda$ were obtained assuming $g^2/4
\pi = 1$ for the new contact interaction coupling. When all
fermions have the same flavor, {\it i.e.\/} $m =n$ in Eq.\
(\ref{lag:4}), we included the normalization $g^2/2 \pi = 1$.
When quarks are running in the loop the color factor ($N_C = 3$)
is taken into account.

In principle, compositeness may not only generate the
four--fermion operators (\ref{lag:4}), which contribute to new
physics at one--loop level, but it may also generate effective
operators which could give tree--level contributions, which we
did not consider here. Thus our bounds are derived under the
assumption that it is unnatural that large cancellations occur
between the tree--level and the one--loop contributions in all
the observables.

\section{Constraints from Precise Measurements}

\subsection{The Leptonic Width of the Z Boson}

The decay rate for the process $Z \rightarrow e^+e^-$ arising from 
the most general vertex expressed in Eq.(\ref{current}) is,
\begin{eqnarray}
\Gamma (Z \to e^+ e^-) &=& \frac{\alpha M_{Z}N_C}{3}\left (1-
\frac{4 m_f^2}{M_Z^2} \right)^{\frac{1}{2}} 
\Biggl [ F_v^2 \left(1+\frac{2 m_f^2}{M_Z^2}\right ) + 
F_a^2 \left(1- \frac{4m_f^2}{M_Z^2}\right )  
\nonumber \\ & &
+ \; 3 F_m F_v + 
F_m^2 \left(1 + \frac{8 m_f^2}{M_Z^2} \right )  \frac{M_Z^2}{8 m_f^2}
+ F_d^2 \left(1 - \frac{4 m_f^2}{M_Z^2} \right )  \frac{M_Z^2}{8 m_f^2}
\Biggr ] \; .
\end{eqnarray}
In this equation, we write the form factors as $F_v = F_v^{tree}
+ \delta F_v$ and $F_a = F_a^{tree} + \delta F_a$, where the
$\delta$'s denote the radiative corrections to the tree level
contribution. When the magnetic and electric dipole contributions
are absent, we get the well known result for the extra 
contribution,
\begin{equation}
\delta \Gamma (Z \to e^+ e^-) = - \; \frac{2 \alpha M_Z N_C}{3}  \; 
\left[ G_V^e \,\, \delta F_v (M_Z^2) -  G_A^e \,\, \delta F_a (M_Z^2)
\right] \; ,
\label{wid}
\end{equation}
where $G_{V,A}^e$ is the vector (axial) couplings of electrons to
the $Z$. Interaction (\ref{lag:4}) also contributes at one--loop
to the width $Z\rightarrow q\bar{q}$ but the effect is too small
for the present experimental accuracy on the hadronic width of
the $Z$. 

The most recent LEP experimental result \cite{lep97},
$\Gamma_{\ell\ell} = 83.91 \pm 0.10$ MeV, can be
compared with the Standard Model predictions for the leptonic
width, in order to
establish bounds on the scale $\Lambda$ through Eq.\ (\ref{wid}).
The Standard Model result depends on the top quark and Higgs
boson masses. In what follows, we present our results for
$m_{\text{top}} = 175$ GeV and for Higgs boson mass $M_H = 300$
GeV, which yield $\Gamma_{\ell\ell} = 83.92$ MeV.

In Table \ref{vec:Z}, we show the 95\% CL bounds on the the scale
of the vector current interactions involving an electron pair and
any other  fermions pair, coming from the measurement of the
leptonic $Z$ width. We should notice that the bounds on several
interactions structures are more restrictive than the ones that
come from direct searches at collider experiments \cite{pdg}.
These cases are denoted in Table \ref{vec:Z} by the numbers in
boldface.

In the case of scalar interactions only the four electron
Lagrangian contributes to the $Z \rightarrow e^+e^-$ width and
leads to the bound $\Lambda \gtrsim 0.6$ TeV for 
$(V_{T}^e V_{T}^i + A_{T}^e A_{T}^i) = 1$, while the
tensorial current interaction does not contribute to this
observable.

\subsection{Anomalous Magnetic and Weak--Magnetic Moment}

The anomalous magnetic form factor is generated only at one--loop
both in the Standard Model and via the four--fermion interaction.
The best determination of the anomalous magnetic moment of the
muon $a^\gamma_\mu \equiv (g_\mu - 2)/2$ comes from a CERN
experiment, {\it i.e.\/} $a^\gamma_\mu = 11 \, 659 \, 230 \, (84)
\times 10^{-10}$ \cite{g-2:exp}. This result should be compared
with the existing theoretical calculations of the QED
\cite{g-2:QED}, electroweak \cite{g-2:EW}, and hadronic
\cite{g-2:had} contributions, which are known with high
precision.

The main theoretical uncertainty comes from the hadronic
contributions which is of the order of $20 \times 10^{-10}$. We
use the limits imposed on $\delta a^{\gamma}_{\mu} \equiv
a^{\gamma}_{\mu} - a^{\gamma}_{\mu(\text{SM})}$ given in Ref.\
\cite{escribano}, $-1.4\times 10^{-8} \leq \delta a^{\gamma}_{\mu}
\leq 2.2 \times 10^{-8}$ at 95\% CL. The proposed AGS experiment
at the Brookhaven National Laboratory \cite{ags} will be able to
measure the anomalous magnetic moment of the muon with an
accuracy of about $\pm 4 \times 10^{-10}$.

The anomalous magnetic moment of electron is measured with great
accuracy $a_e^{\gamma} = 1\;159\;652\;188.4 (4.3) \times
10^{-12}$ \cite{aeexp}. From the comparison of this result with
the theoretical value \cite{kinoshita} the authors of Ref.\
\cite{escribano} find the limits: $-6.9 \times 10^{-11}
\leq \delta a_e^{\gamma} \leq 4.3 \times 10^{-11} $ with 95\% CL.
We use these limits in our calculations.

We present in Table \ref{mag:mom} the bounds on $\Lambda$ from
the present $g - 2$ data and also from the forthcoming AGS
experiment, for the tensorial current interaction assuming
$(V_{T}^e V_{T}^i - A_{T}^e A_{T}^i) = 1$.  For the scalar
current interaction involving four identical leptons,  the bounds
on $\Lambda$ reads: $\Lambda \gtrsim 0.1 (1.0)$ TeV for electron
(muon) and  $\Lambda \gtrsim 7$ TeV for the AGS muon experiment
for $(V^u_S V^l_S - A^u_S A^l_S) = 1$.

Bernabeu {\sl et al.\/} \cite{berna} evaluated the Standard Model 
contribution to the anomalous weak--magnetic moment and discuss
the possibility of its measurement through the analysis of the
angular asymmetry of the semileptonic $\tau$--lepton decay products.
Assuming  that the $\tau$ direction is fully reconstructed, they
obtain  a sensitivity of the order of $|a^Z_\tau (M_Z^2)|
\lesssim 10^{-4}$. Present limits limits on the four--fermion
interactions  strongly reduces the possibility of  observing its
effect on the anomalous weak--magnetic moment of the $\tau$
lepton at this level of sensitivity. 

\subsection{Electric and Weak Dipole Moment}

A nonzero electric dipole moment for the leptons is forbidden by
both $T$ and $P$ invariance. The best present bound on the
electric dipole moment of the electron comes from its measurement
using the atomic--beam magnetic resonance method with the
$^{205}$Tl atom and reads $|d_e| \le 4.4 \times 10^{-27}$ e cm
\cite{d:e}. For the muon, the present limit on electric dipole
moment is $|d_\mu| \le 9.3  \times 10^{-19}$ e cm \cite{d:mu}.
For the $\tau$ lepton the limits on the real and imaginary
parts of the weak dipole moment were measured to be \cite{d:tau}
$|\text{Re}(d^w_{\tau})| \le 4.8  \times 10^{-18}$ e cm, and
$|\text{Im}(d^w_{\tau})| \le 0.93 \times 10^{-17}$ e cm.

In Table \ref{dip}, we show the bound on the scale of the
tensorial current interaction from the limits on the electron and
muon electric dipole moment for $(A_{T}^e V_{T}^i + A_{T}^i
V_{T}^e) = 1$. For the $\tau$--lepton, we can establish a limit on
this interaction from the real part of the weak dipole moment
measured at the $Z$ from the reconstruction of $\tau^+ \tau^-$
events given above. A bound of $\Lambda \gtrsim 260$
GeV is obtained when a top quark of $m_{\text{top}} = 175$ GeV
is running in the loop. 

For the scalar contact interaction of four equal leptons, we get
$\Lambda \gtrsim 45$ TeV for electrons and just $\gtrsim 25$ GeV
for muons, assuming that $(V^l_S A^u_S + V^u_S A^l_S) = 1$.

\section{Conclusions}

In this paper we have evaluated the one--loop contribution to the
leptonic  $Z$ width, anomalous magnetic, weak--magnetic, electric
and weak--dipole moments of leptons arising from the four-Fermi
Lagrangian.  Using the precise measurements of these parameters
we extract bounds on the energy scale of these effective
interactions.

Our results show that vector and axial vector interactions
involving leptons in different combinations can be strongly
constrained by the $Z$--width measurement. In particular, our
bounds on four--Fermi interactions, for some current structures,
are more restrictive than the ones from direct search at collider
experiments. Scalar and tensor interactions can also be
constrained from the measurement of the anomalous magnetic and
weak-magnetic moments and from dipole and weak-dipole moments.
Our results show that for interactions involving electrons the
strongest bounds are derived from the precision measurement of
the dipole electric moment of the electron, while for
interactions involving muons the best bounds arise from their
contributions to the anomalous magnetic moment of the muon. 

\acknowledgments
M.\ C.\ Gonzalez--Garcia is grateful to the Instituto de
F\'{\i}sica Te\'orica of the Universidade  Estadual Paulista for
its kind hospitality. This work was supported by Funda\c{c}\~ao
de Amparo \`a Pesquisa do Estado de S\~ao Paulo, by DGICYT under
grant PB95-1077, by CICYT under  grant AEN96-1718, and by
Conselho Nacional de Desenvolvimento Cient\'{\i}fico e
Tecnol\'ogico.


\protect

\protect

%
%

\begin{table}
\caption{95 \% CL bound on $\Lambda$, in TeV, from the leptonic
width of $Z$ for different chiralities of the vector couplings.}
\label{vec:Z}
\begin{tabular}{||c|c|c|c|c|c|c|c|c|c||}
Channel & $\Lambda_{\eta}$ & LL & RR & LR & RL & VV & AA & LL+RR & LR+RL \\
\tableline \tableline
$ee$ & $\Lambda_{+}$ & 1.1 & 0.56 & 0.7 & 0.7 & 1.4 & 2.2 & 1.8 & 1.1
\\[-0.12cm]
& $\Lambda_{-}$ & 1.2 & 0.65 & 0.6 & 0.6 & 1.6 & 2.5 & 2.0 & 1.0 \\
\tableline
$\mu \mu$ &$\Lambda_{+}$ & 1.1 & 0.89 & 1.1 & 1.1 & --- & 2.3 & 1.5 & 1.7 
\\[-0.12cm]
&$\Lambda_{-}$ & 1.2 & 1.0 & 1.0 & 1.0 & --- & 2.6 & 1.7 & 1.5 \\
\tableline
$\tau \tau$ &$\Lambda_{+}$ & 1.1 & 0.89 & 1.1 & 1.1 & --- & 2.3 & 1.5 & 
1.7 \\[-0.12cm]
&$\Lambda_{-}$ & 1.2 & 1.0 & 1.0 & 1.0 & --- & 2.6 & {\bf 1.7} & 1.5 \\
\tableline
$\ell \ell$ \tablenotemark[1] &$\Lambda_{+}$ & 2.5  & 1.6 & 1.9 & 1.9 & 
1.4 & 4.3 & 3.1 &  2.9 \\[-0.12cm]
&$\Lambda_{-}$ & 2.8  & 1.8 & 1.7 & 1.7 & 1.6 & 4.8 & 3.5 & 2.6 \\
\tableline
$uu$ & $\Lambda_{+}$ & {\bf 2.7}  & {\bf 1.6} & {\bf 2.2} & 1.5 & 0.5 & 
{\bf 4.9} & 
3.3 & {\bf 2.9} \\[-0.12cm]
&$\Lambda_{-}$ & 2.4  & 1.4 & {\bf 2.5} & 1.7 & 0.5 & {\bf 4.3} & 2.9 & 
{\bf 3.2} \\
\tableline
$dd$ & $\Lambda_{+}$ & 2.7  & 0.9 & {\bf 2.8} & 1.1 & 0.7 & {\bf 4.3} & 
3.0 & {\bf 3.2} \\
&$\Lambda_{-}$ & {\bf 3.0} & 1.0 & {\bf 2.5} & 1.0 & 0.8 & {\bf 4.9} & 
{\bf 3.4} &
 {\bf 2.8} \\
\tableline
$(uu + dd)$ 
& $\Lambda_{+}$ & 1.0 & 1.0 & 1.0 & 1.0 & 0.4 & ---  & 0.2 & 0.2  \\[-0.12cm]
& $\Lambda_{-}$ & 1.1 & 0.9 & 0.9 & 1.1 & 0.5 & ---  & 0.3 & 0.3 \\
\tableline
$cc$ 
& $\Lambda_{+}$ & 2.7 & 1.6 & 2.2 & 1.5 & 0.5 & 4.9 & 3.3  & 2.9 \\[-0.12cm]
& $\Lambda_{-}$ & 2.4 & 1.4  & 2.5 & 1.7 & 0.5 & 4.3 & 2.9  & 3.2  \\
\tableline
$bb$ & $\Lambda_{+}$ & 2.7  & 0.9 & {\bf 2.8} & 1.1 & 0.7 & 4.3 & 
3.0 & {\bf 3.1} \\[-0.12cm]
&$\Lambda_{-}$ & {\bf 3.1} & 1.0 & {\bf 2.5} & 0.9 & 0.8 & {\bf 4.8} & 3.3 & 
2.8 \\
\tableline
$tt$ \tablenotemark[2]& $\Lambda_{+}$ & 11.5 & 10.7 & 11.8 & 13.0 & 0.5 & 
23.9 & 
16.3 & 18.2 \\[-0.12cm]
&$\Lambda_{-}$ & 12.7 & 12.0 & 10.5 & 11.6 & 0.5 & 26.7 & 18.2 & 16.3 \\
\tableline
$qq$ \tablenotemark[3] & $\Lambda_{+}$ & 3.3 & 1.0 & {\bf 3.4} & 
1.0 & 1.1 & 4.3 & 3.0 & {\bf 3.0} \\[-0.12cm]
&$\Lambda_{-}$ & {\bf 3.7}   & 0.9 & {\bf 3.0} & 1.1 & 1.2 & {\bf 4.8} & 
{\bf 3.4} & 2.7 \\
\end{tabular}
\tablenotetext[1]{$\ell'\ell' = ee + \mu\mu + 
\tau\tau$.}
\tablenotetext[2]{$m_{\text{top}}=175$ GeV.}
\tablenotetext[3]{$qq = uu + dd + ss + cc + bb$.}
\end{table}

\begin{table}
\caption{95 \% CL bound on the scale $\Lambda$ of the tensorial current 
interaction, in TeV, from the electron and muon anomalous magnetic moments.} 
\label{mag:mom}
\begin{tabular}{||c|c|c|c||}
Channel & $\ell =$ electron &  $\ell =$ muon (present) &  $\ell =$
 muon (AGS)    \\
\tableline\tableline
& $\eta = +1$ / $\eta = -1$ & $\eta = +1$ / $\eta = -1$ & \\
\tableline
$\ell\ell\ell\ell $  & 0.40/0.31 & 3.2/4.0 & 26 \\
\tableline
$ee\mu\mu$ 	& 7.3/5.6 & 0.22/0.28 & 1.8 \\
\tableline
$ee\tau\tau$  	& 31/24 & --- & ---   \\
\tableline
$\mu\mu\tau\tau$  & --- & 16/20 & 130   \\
\tableline
$\ell'\ell'\ell\ell$ \tablenotemark[1] & 32/25 & 17/21 & 133  \\
\tableline
$uu\ell\ell$ 	& 1.7/2.1 & 1.4/1.1 & 8.7   \\
\tableline
$dd\ell\ell$ 	& 2.1/1.7 & 1.1/1.4 & 8.7   \\
\tableline
$ss\ell\ell$ 	& 10.1/8.0 & 5.1/6.5 & 42   \\
\tableline
$cc\ell\ell$ 	& 32/41 & 27/21 & 169   \\
\tableline
$bb\ell\ell$ 	& 53/42 & 28/35 & 221   \\
\tableline
$tt\ell\ell$ \tablenotemark[2] 	& 367/468 & 317/251 & 1989  \\
\tableline
$qq\ell\ell$ \tablenotemark[3] & 35/27 & 18/23 & 144  \\
\end{tabular}
\tablenotetext[1]{$\ell'\ell' = ee + \mu\mu + 
\tau\tau$.}
\tablenotetext[2]{$m_{\text{top}}=175$ GeV.}
\tablenotetext[3]{$qq = uu + dd + ss + cc + bb$.}
\end{table}

\begin{table}
\caption{95\% CL bound on scale $\Lambda$ of the tensorial
current interaction  from the electric dipolar moment. Note the
different unities for electrons and muons}
\label{dip}
\begin{tabular}{||c|c|c||}
Channel & $\ell =$ electron (TeV) & $\ell =$ muon (GeV)    \\
\tableline
$\ell\ell\ell\ell$  	& 132 & 77  \\
\tableline
$ee\mu\mu$ 		& 2053 & 3.9   \\
\tableline
$ee\tau\tau$  		& 8670 & ---  \\
\tableline
$\mu\mu\tau\tau$ 	& --- & 348   \\
\tableline
$\ell'\ell'\ell\ell$ \tablenotemark[1] & 8800 & 359 \\
\tableline
$uu\ell\ell$    	& 613 & 21 \\
\tableline
$dd\ell\ell$    	& 613 & 21  \\
\tableline
$ss\ell\ell$ 		& 2842 & 108    \\
\tableline
$cc\ell\ell$ 		& 11340 & 460    \\
\tableline
$bb\ell\ell$ 		& 14719 & 604    \\
\tableline
$tt\ell\ell$ \tablenotemark[2]& 128573 & 5669  \\
\tableline
$qq\ell\ell$ \tablenotemark[3] & 9750 & 390  \\
\end{tabular}
\tablenotetext[1]{$\ell'\ell' = ee + \mu\mu + 
\tau\tau$.}
\tablenotetext[2]{$m_{\text{top}}=175$ GeV.}
\tablenotetext[3]{$qq = uu + dd + ss + cc + bb$.}
\end{table}


\begin{references}
\bibitem{compo} M.\ E.\ Peskin, Proceedings of the 1985
International Symposium on Lepton and Photon Interactions at High
Energies, Kyoto, 1985, p.\ 714,  eds. M.\ Konuma and K.\
Takahashi; E.\ Eichten, K.\ Lane, and M.\ Peskin, Phys.\ Rev.\
Lett.\ {\bf 50} 811 (1983).

\bibitem{four}  G.\ Altarelli, J.\ Ellis, G.\ F.\ Giudice, S.\
Lola, M.\ L.\ Mangano  Nucl.\ Phys.\ {\bf B506} 3, (1997); 
V.\ Barger, K.\ Cheung, K.\ Hagiwara and D.\ Zeppenfeld, Phys.\
Rev.\ {\bf D57}, 391 (1998);
Gi--C.\ Cho, K.\ Hagiwara and  S.\ Matsumoto, preprint
KEK-TH-521, and hep--ph/9707334; 
K.\ Akama, K.\ Katsuura  Phys.\ Rev.\ {\bf D56} 2490, (1997); 
N.\ G.\ Deshpande, B.\ Dutta, X.\ He  Phys.\ Lett.\  {\bf B408}
288, (1997);
N.\ Bartolomeo, M.\ Fabbrichesi, Phys.\ Lett.\ {\bf B406} 237 (1997). 

\bibitem{our:4} M.\ C.\ Gonzalez--Garcia and S.\ F.\ Novaes,
Phys.\ Lett.\ {\bf B407}, 255 (1997).

\bibitem{hera} C.\ Adloff {\it et al.} (H1 Collaboration), Z.\
Phys.\ {\bf C74}, 191 (1997); J.\ Breitweg {\it et al.} (ZEUS
Collaboration),  Z.\ Phys.\ {\bf C74}, 207 (1997).

\bibitem{spiesberger} R.\ Ruckl and H.\ Spiesberger, hep-ph/9710327.

\bibitem{tevatron} F.\ Abe {\it et al.} (CDF Collaboration),
Phys.\ Rev.\ Lett.\ {\bf 77}, 438 (1996), {\it idem} {\bf 77},
5336 (1996), {\it idem} {\bf 79}, 2198 (1997); B.\ Abbott {\it et
al.} (D\O ~Collaboration), preprint FERMILAB-PUB-97-237-E and
hep-ex/9707016.

\bibitem{reg:dim} G.\ 't Hooft and M.\ Veltman, Nucl.\ Phys.\
{\bf B44}, 189 (1972); C.\ G.\ Bollini and J.\ J.\ Giambiagi,
Nuovo Cim.\ {\bf 12B}, 20 (1972). 

\bibitem{feyn} R.\ Mertig, M.\ Bohm and A.\ Denner, Comput.\
Phys.\ Commun.\ {\bf 64}, 345 (1991).

\bibitem{ward} J.\ C.\ Ward, Phys.\ Rev.\ {\bf 78} 182 (1950);
Y.\ Takahashi, Nuovo Cim.\ {\bf 6} 317 (1957); J.\ C.\ Taylor,
Nucl.\ Phys.\ {\bf B33} 436 (1971); A.\ A.\ Slavnov, Theor.\ and
Math.\ Phys.\ {\bf 10} 99 (1972).

\bibitem{lep97} The LEP Collaborations ALEPH, DELPHI, L3, OPAL,
the LEP Electroweak Working Group and the SLD Heavy Flavour
Group, from  the Contributions of the LEPand SLD experiments to
the 1997 summer conferences,  LEPEWWG/97-02 (1997).

\bibitem{pdg} R.\ M.\ Barnett {\it et al.}, Phys.\ Rev.\ D54, 1
(1996) and 1997 off--year partial update for the 1998 edition
available on the PDG WWW pages (URL: \verb+http://pdg.lbl.gov/+).

\bibitem{g-2:exp} J.\ Bailey {\it et al.}, Nucl.\ Phys.\ {\bf
B150}, 1 (1979); E.\ R.\ Cohen and B.\ N.\ Taylor, Rev.\ Mod.\
Phys.\ {\bf 59}, 1121 (1987).

\bibitem{g-2:QED} For a review see: T.\ Kinoshita and W.\ J.\
Marciano, ``Quantum Electrodynamics'', edited by T.\ Kinoshita,
World Scientific, Singapore (1990), p.\ 419 and references
therein.

\bibitem{g-2:EW}
G.\ Altarelli, N.\ Cabibbo and L.\ Maiani, Phys.\ Lett.\ {\bf
B40}, 415 (1972); 
R.\ Jackiw and S.\ Weinberg, Phys.\ Rev.\ D{\bf 5}, 2473 (1972);
I.\ Bars and M.\ Yoshimura, Phys.\ Rev.\ D{\bf 6}, 374 (1972);
K.\ Fujikawa, B.\ W.\  Lee and A.\ I.\ Sanda, Phys.\ Rev.\ D{\bf 6}, 
2923 (1972); 
W.\ A.\  Bardeen,  R.\ Gastmans and B.\ E.\ Lautrup,  Nucl.\
Phys.\ {\bf B46}, 319 (1972); 
T.\ V.\  Kukhto, E.\ A.\  Kuraev, A.\ Schiller and Z.\ K.\
Silagadze, Nucl.\ Phys.\ {\bf B371}, 567 (1992);
S.\  Peris, M.\  Perrottet and E.\ de Rafael, Phys.\ Lett.\ {\bf
B355}, 523 (1995);
A.\ Czarnecki, B.\ Krause and W.\ J.\ Marciano, Phys.\ Rev.\ D{\bf
52}, 2619 (1995), and Phys.\ Rev.\ Lett.\ {\bf 76}, 3267 (1996).

\bibitem{g-2:had}
J.\ Calmet, S.\ Narison, M.\ Perrottet and E.\ de Rafael, Phys.\
Lett.\ {\bf B61}, 283 (1976); {\it idem} Rev.\ Mod.\ Phys.\ {\bf
49}, 21 (1977); 
T.\ Kinoshita, B.\ Ni{\u{z}}i\'c and Y.\ Okamoto, 
E.\ de Rafael, Phys.\ Lett.\ {\bf B322}, 239 (1994);
E.\ Pallante, Phys.\ Lett.\ {\bf B341}, 221 (1994);
M.\ Hayakawa, T.\ Kinoshita and  A.\ I.\ Sanda, Phys.\ Rev.\ 
Lett.\ {\bf 75}, 790 (1995);
J.\ Bijnens, E.\ Pallante and J.\ Prades, Phys.\ Rev.\ Lett.\ {\bf
75}, 1447 (1995); 
S.\ Eidelman and F.\ Jegerlehner, Z.\ Phys.\ {\bf C67}, 585 (1995);
K.\ Adel and F.\ J.\ Yndur\'ain, Univ. Aut\'onoma de Madrid
preprint, FTUAM 95--32, hep--ph/9509378 (1995).

\bibitem{escribano} R.\ Escribano and E.\ Masso, hep-ph/9607218.

\bibitem{ags}
B.\ L.\ Roberts, Z.\ Phys.\ {\bf C56}, S101 (1992).

\bibitem{aeexp} R.\ S.\ Van Dyck, Jr., P.\ B.\ Schwinberg and H.\
G.\ Dehmelt, Phys.\ Rev.\ Lett.\ {\bf 59}, 26 (1987).

\bibitem{kinoshita} T.\ Kinoshita, Phys.\ Rev.\ Lett.\ {\bf 75}, 
4728 (1995).

\bibitem{berna}
J.\ Bernabeu, G.\ A.\ Gonzalez--Sprinberg, M.\ Tung and  J.\
Vidal, Nucl.\ Phys.\ {\bf B436}, 474 (1995).

\bibitem{d:e} E.\ D.\ Commins, S.\ B.\ Ross, D.\ DeMille and B.\ C.\ 
Regan, Phys.\ Rev.\ A{\bf 50}, 2960 (1994).

\bibitem{d:mu} J.\ Bailey {\it et al.} (CERN Muon Storage Ring
Collaboration),  J.\ Phys.\ {\bf G4}, 345 (1978).

\bibitem{d:tau} K.\ Ackerstaff {\it et al.} (Opal Collaboration), 
Z.\ Phys.\, {\bf C74}, 403 (1997).

\end{references}
\end{document}